\documentclass[aps,pra,floatfix,twocolumn,superscriptaddress,showpacs,10pt]{revtex4-1}
\usepackage{amssymb, amsmath,color,mciteplus,graphicx,subfigure}
\usepackage{calc,epsfig,epstopdf,color,mciteplus,bm,mathrsfs}
\usepackage{times}
\usepackage{float}
\usepackage{lipsum}

\begin{document}

\title{Error-Tolerant Geometric Quantum Control for Logical Qubits with Minimal Resource}

\author{Tao Chen}
\affiliation{Guangdong-Hong Kong Joint Laboratory of Quantum Matter, Department of Physics, and HKU-UCAS Joint Institute for Theoretical and Computational Physics at Hong Kong,\\ The University of Hong Kong, Pokfulam Road, Hong Kong, China}
\affiliation{Guangdong Provincial Key Laboratory of Quantum Engineering and Quantum Materials, 
and School of Physics\\ and Telecommunication Engineering, South China Normal University, Guangzhou 510006, China}

\author{Zheng-Yuan Xue}\email{zyxue83@163.com}
\affiliation{Guangdong Provincial Key Laboratory of Quantum Engineering and Quantum Materials, 
and School of Physics\\ and Telecommunication Engineering, South China Normal University, Guangzhou 510006, China}
\affiliation{Guangdong-Hong Kong Joint Laboratory of Quantum Matter,  and Frontier Research Institute for Physics,\\ South China Normal University, Guangzhou 510006, China}

\author{Z. D. Wang}\email{zwang@hku.hk}
\affiliation{Guangdong-Hong Kong Joint Laboratory of Quantum Matter, Department of Physics, and HKU-UCAS Joint Institute for Theoretical and Computational Physics at Hong Kong,\\ The University of Hong Kong, Pokfulam Road, Hong Kong, China}
\affiliation{Guangdong-Hong Kong Joint Laboratory of Quantum Matter,  and Frontier Research Institute for Physics,\\ South China Normal University, Guangzhou 510006, China}

\date{\today}

\begin{abstract}
  Geometric quantum computation offers a practical strategy toward robust quantum computation due to its inherently error tolerance.  However, the rigorous geometric conditions lead to complex and/or error-disturbed quantum controls, especially for logical qubits that involve more physical qubits, whose error tolerance is effective in principle though, their experimental demonstration is still demanding. Thus, how to best simplify the needed control and manifest its full advantage has become the key to widespread applications of geometric quantum computation. Here we propose a new fast and robust geometric scheme, with the decoherence-free-subspace encoding, and present its physical implementation on superconducting quantum circuits, where we only utilize the experimentally demonstrated parametrically tunable coupling to achieve high-fidelity geometric control over logical qubits. Numerical simulation verifies that it can efficiently combine the error tolerance from both the geometric phase and logical-qubit encoding, displaying our gate-performance superiority over the conventional dynamical one without encoding, in terms of both gate fidelity and robustness. Therefore, our scheme can consolidate both error suppression methods for logical-qubit control, which sheds light on the future large-scale quantum computation.
\end{abstract}

\maketitle

\section{Introduction}

The superiority of quantum computation over classical one essentially lies in the superposition and entanglement of  quantum systems \cite{QC}. But an inescapable truth is that, the effects of the environment-induced decoherence and inaccurate control on a quantum system will greatly limit the performance of quantum gates, which are the building blocks for executing quantum computation, thus revealing the necessity for exploring a method to combat both decoherence and control errors, especially for future large-scale quantum circuits. Geometric phases \cite{BerryP,non-Abelian,A-Aphase,noncyclicPphase}, due to their inherent error-tolerant feature, have become critical elements for realizing high-fidelity and robust quantum controls. However, for adiabatic geometric quantum computation (GQC) \cite{HQC1,HQC2,NMRAGQC,Duan,SQAGQC}, due to long evolution time needed to satisfy the adiabatic conditions, the resulting gate fidelities are relatively low, despite they are indeed being more robust against control errors.

Recently, GQC  \cite{NGp,NGQC,UGQC,NHQC,NGQCSL,NGQCSLSQ,  DGQC} based on the  nonadiabatic geometric phases \cite{A-Aphase,noncyclicPphase} have been proposed to implement robust and high-fidelity quantum gates, which eliminate the restriction of slow evolution. Remarkably, experimental demonstrations for elementary geometric quantum gates have also been achieved on various systems, such as trapped ions \cite{Geo-ion1,Geo-ion2}, NMR \cite{Geo-NMR1,Geo-NMR2,Geo-NMR3,Geo-NMR4}, superconducting quantum circuits \cite{Geo-SQ1,Geo-SQ2,Geo-SQ3,Geo-SQ4,Geo-SQ5,Geo-SQ6,Geo-SQ7}, and nitrogen vacancy centers \cite{Geo-NV1,Geo-NV2,Geo-NV3,Geo-NV4,Geo-NV5,Geo-NV6}, etc. Meanwhile, to further consolidate the geometric robustness, many efforts have been made to make GQC being compatible with various optimal-control techniques, including the composite pulse \cite{NGQC+Com1,NGQC+Com2}, dynamical decoupling \cite{NGQC+DD1,NGQC+DD2}, time-optimal control \cite{NGQC+TOC}, path optimization \cite{NGQC-Path}, etc.  Nonetheless, the rigorous geometric conditions  lead to complex and error-disturbed quantum controls, so that  existing  GQC is favourable merely for a certain kind of control error.  Thus, it is still quite challenging to demonstrate the full error-tolerant advantages for GQC.

Besides the control errors, the environmental induced decoherence is another central obstacle in realizing GQC. To this end, the proposed geometric schemes \cite{NHQC-DFS1,NHQC-DFSion,NHQC-DFSatomC, NHQC-DFS2,NHQC-NoiseC,DFSSQ,UGQC-DFS,NHQC-2qubitDFS} under the protection of decoherence-free-subspace (DFS) encoding \cite{DFS1,DFS2,DFS3,DFS4} aim to achieve the combination of the geometric error-tolerant features and decoherence resilience of the encoding. However, due to more involved physical qubits, more error sources are introduced too, such that the logical-qubit manipulation becomes complicated and  fail to achieve better gate performance under their hybrid protection in subsequent experimental demonstration.

Here we aim to solve the two aforementioned key problems in GQC: (i) to best simplify the geometric quantum control that manifests full advantage of GQC; (ii) to make GQC more compatible with the protection of DFS encoding. To this end, from a general approach to construct geometric quantum gate, we propose a new geometric scheme that can shorten the operation time and possess the robustness advantage beyond the conventional dynamical scheme. In addition, with the pursuit of a minimal number of physical qubits, we merely utilize parametrically tunable coupling between a transmon and a microwave resonator to achieve high-fidelity geometric control for DFS logical qubits on superconducting quantum circuit. Our scheme does not need additional  auxiliary level/qubit, and thus avoids introducing more error sources. Numerical simulation verifies that our scheme can fully consolidate the hybrid-protection capabilities from both geometric phase and the encoding, by displaying absolute error-tolerant advantages over the dynamical gates without encoding, in terms of both frequency-drift error and decoherence, which are the main error source for superconducting quantum circuits. Therefore, our scheme shows the prospect of GQC with logical-qubit encoding that is inevitable in future fault-tolerant quantum computation.

\begin{figure}[tbp]
  \centering
  \includegraphics[width=0.9\linewidth]{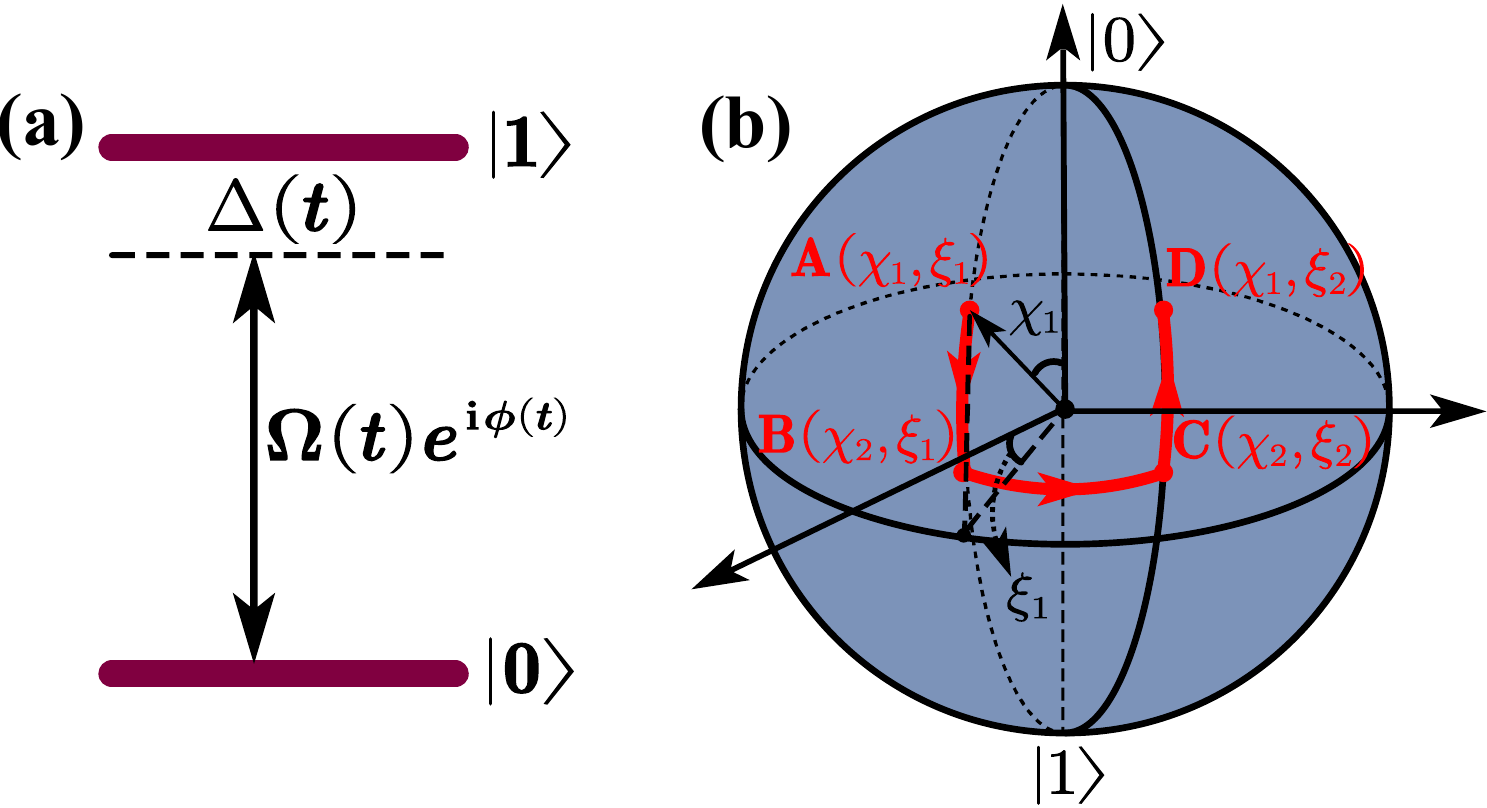}
  \caption{(a) A simple two-level coupling structure driven by a microwave field. (b) Illustration of the evolution path $\textbf{A}\!\!\rightarrow \!\!\textbf{B} \!\!\rightarrow \!\!\textbf{C}\!\!\rightarrow \!\!\textbf{D}$ of state $|\Psi_0(t)\rangle$ on a Bloch sphere.}
  \label{BlochSphere}
\end{figure}

\section{A general scheme for geometric gates}

We first consider a general two-level quantum system that consists of a ground state $|0\rangle=(1,0)^\dag$ and an excited state $|1\rangle=(0,1)^\dag$, as shown in Fig. \ref{BlochSphere}(a). Its arbitrary quantum control can be realized by a microwave drive, as described by the Hamiltonian of
\begin{eqnarray} \label{H}
\mathcal{H}(t)=\frac{1}{2}\textbf{B}(t)\cdot \bm{\sigma},
\end{eqnarray}
where control field $\textbf{B}(t)$ has three directions, which are $B_x=\Omega(t)\cos\phi(t)$, $B_y=\Omega(t)\sin\phi(t)$ and $B_z=-\Delta(t)$, respectively; $\bm{\sigma}=(\sigma_x, \sigma_y, \sigma_z)$ are the Pauli matrices; $\Omega(t)$ and $\phi(t)$ are the time-dependent driving amplitude and phase of the microwave field, and $\Delta(t)$ is the time-dependent detuning. There is a pair of orthogonal evolution states $|\Psi_i(t)\rangle=e^{\textrm{i}f_i(t)}|\psi_i(t)\rangle$ with subscript $i\!=\!0,1$, $|\psi_0(t)\rangle\!=\!\cos{\frac {\chi(t)} {2}}|0\rangle\!+\sin{\frac {\chi(t)} {2}}e^{\textrm{i}\xi(t)}|1\rangle$ and $|\psi_1(t)\rangle\!=\!\sin{\frac {\chi(t)} {2}}e^{-\textrm{i}\xi(t)}|0\rangle-\cos{\frac {\chi(t)} {2}}|1\rangle$, where the phase factor $f_i(t)$ is a global phase of $|\Psi_i(t)\rangle$, $\chi(t)$ and $\xi(t)$ represent the time-dependent change of polar and azimuth angles of the state vector on a Bloch sphere, as shown in Fig. \ref{BlochSphere}(b). Utilizing the Schr\"{o}dinger equation as $\textrm{i}|\dot{\Psi}_i(t)\rangle=\mathcal{H}(t)|\Psi_i(t)\rangle$, we solve for
\begin{eqnarray} \label{paralimt1}
\dot{\chi}(t)&=&-B_x\sin\xi(t)+B_y\cos\xi(t), \notag\\
\dot{\xi}(t)&=&-\Delta(t)-\cot\chi(t)\left[B_x\cos\xi(t)+B_y\sin\xi(t)\right],
\end{eqnarray}
thereby determining the dependence of evolution details of state $|\Psi_i(t)\rangle$ and the control parameters of $\mathcal{H}(t)$. Besides, after a time period $\tau$, accumulated global phase $f_0(\tau)=-f_1(\tau)=\gamma$ includes the dynamical part $\gamma_d=\!-\!\int^\tau_0\langle \Psi_i(t)|\mathcal{H}(t)|\Psi_i(t)\rangle \textrm{d}t$ and the geometric part
\begin{eqnarray} \label{GeoP}
\gamma_g=\gamma-\gamma_d=-\frac {1} {2}\int^\tau_0 \dot{\xi}(t)[1-\cos\chi(t)] \textrm{d}t,
\end{eqnarray}
where the geometric nature \cite{A-Aphase,noncyclicPphase} of $\gamma_g$ comes from the fact that it is given by half of the solid angle enclosed by the noncyclic evolution path and its geodesic connecting the initial point [$\chi(0), \xi(0)$] and the final point [$\chi(\tau), \xi(\tau)$].
In this way, we can determine the pure geometric property of the evolution process by designing state parameters $\dot{\xi}(t)\sin^2\!\chi(t)\!=\!-\Delta(t)$ to meet $\gamma_d\!=\!0$. Thus, submitting this geometric condition into Eq. (\ref{paralimt1}), we can use the following relations
\begin{eqnarray} \label{paralimt2}
B_x&=&-\dot{\xi}(t)\sin\chi(t)\cos\chi(t)\cos\xi(t)-\dot{\chi}(t)\sin\xi(t), \notag\\
B_y&=&-\dot{\xi}(t)\sin\chi(t)\cos\chi(t)\sin\xi(t)+\dot{\chi}(t)\cos\xi(t), \notag\\
B_z&=&-\dot{\xi}(t)\sin^2\!\chi(t),
\end{eqnarray}
to reversely fix the control parameters of Hamiltonian $\mathcal{H}(t)$.

After an evolution period $\tau$, two evolution states undergo a change as $|\Psi_i(0)\rangle\!\rightarrow\!|\Psi_i(\tau)\rangle\!= \!e^{(-1)^i\textrm{i}\gamma_g}|\psi_i(t)\rangle$, the associated evolution operator
\begin{eqnarray} \label{U}
U(\tau,0)\!=\!e^{\textrm{i}\gamma}|\psi_0(\tau)\rangle\langle \psi_0(0)|\!+\!e^{-\textrm{i}\gamma}|\psi_1(\tau)\rangle\langle \psi_1(0)| \quad\quad\quad\quad \\
=\!\left(\!\!
\begin{array}{cccc}
(c_{\gamma'}c_{\chi_-} \!+\!\textrm{i}s_{\gamma'} c_{\chi_+}) e^{-\textrm{i} \xi_-}  & (-c_{\gamma'}s_{\chi_-}\!+\!\textrm{i}s_{\gamma'} s_{\chi_+}) e^{-\textrm{i} \xi_+} \\
(c_{\gamma'}s_{\chi_-} \!+\!\textrm{i}s_{\gamma'} s_{\chi_+}) e^{\textrm{i} \xi_+} & (c_{\gamma'}c_{\chi_-} \!-\!\textrm{i}s_{\gamma'} c_{\chi_+}) e^{\textrm{i} \xi_-}
\end{array}
\!\!\right),\notag
\end{eqnarray}
can be obtained, where $c_j\!=\!\cos j$, $s_j\!=\!\sin j$, $p_\pm=[p(\tau)\pm p(0)]/2$ and $\gamma'=\gamma_g+\xi_-$. Thus, arbitrary geometric gates are realized by setting boundary value of $\chi(t)$ and $\xi(t)$.

\section{A new construction of geometric gates}

In the above general geometric framework, we can also set time-dependent shapes of parameters $\chi(t)$ and $\xi(t)$ to realize different geometric evolution processes. We here design a new geometric evolution scheme with state parameters $(\chi(t), \xi(t))$ as follows:
\begin{eqnarray} \label{path}
\chi(0)&\!=\!&\chi_1 \bm{\rightsquigarrow} \chi(\tau_1)\!=\!\chi_2 \bm{\rightarrow}  \!\chi(\tau_2)\!=\!\chi_2 \bm{\rightsquigarrow} \chi(\tau)=\!\chi_1 \notag\\
\xi(0)&\!=\!&\xi_1 \bm{\rightarrow} \xi(\tau_1)=\xi_1 \bm{\rightsquigarrow} \xi(\tau_2)=\xi_2 \bm{\rightarrow} \xi(\tau)=\xi_2
\end{eqnarray}
where arrows ``$\bm{\rightarrow}$" and ``$\bm{\rightsquigarrow}$" indicate whether the parameters $\chi(t)$ and $\xi(t)$ remain constant or change with time during the evolution process, respectively. For example, in the first time segment $t\in[0,\tau_1]$, $\chi(t)$ changes from $\chi_1$ to $\chi_2$, while $\xi(t)=\xi_1$ remains constant. The visualized state-evolution details on a Bloch sphere are shown in Fig. \ref{BlochSphere}(b), that is: start from point \textbf{A}$(\chi_1, \xi_1)$ and evolve along the longitude line with $\xi(t)=\xi_1$ to point \textbf{B}$(\chi_2, \!\xi_1)$ at the time $\tau_1$; then evolve along the latitude line with $\chi(t)=\chi_2$ to point \textbf{C}$(\chi_2, \xi_2)$ at the time $\tau_2$; finally evolve to point \textbf{D}$(\chi_1, \xi_2)$ along the longitude line with $\xi(t)\!=\!\xi_2$ at the final time $\tau$. Therefore, under the above evolution process, according to the parameter-limited relation in Eq. (\ref{paralimt2}), we can determine the Hamiltonian parameters $\Omega(t)$ and $\phi(t)$ in these three segments $t\in[0, \tau_1]$, $[\tau_1, \tau_2]$ and $[\tau_2, \tau]$ as
\begin{eqnarray} \label{3path}
\int^{\tau_1}_0\Omega(t)\textrm{d}t&=&|\chi_2-\chi_1|, \quad\quad\quad\quad\ \ \phi(t)=\xi_1+\frac{\pi}{2}, \notag\\
\int^{\tau_2}_{\tau_1}\Omega(t)\textrm{d}t&=&2(\xi_2-\xi_1)\sin(2\chi_2), \ \ \phi(t)=\xi(t)+\pi, \notag\\
\int^{\tau}_{\tau_2}\Omega(t)\textrm{d}t&=&|\chi_2-\chi_1|, \quad\quad\quad\quad\ \ \phi(t)=\xi_2-\frac{\pi}{2},
\end{eqnarray}
respectively, with detuning $\Delta(t)=0, -\Omega(t)\tan\chi_2, 0$, where $\xi(t)\!=\!\xi_1\!+\!\int^{t}_{\tau_1}\!\!\Omega(t')\textrm{d}t'\!/[2\sin(2\chi_2)]$. This is the case of $\chi_2>\chi_1$, and when $\chi_2<\chi_1$, parameter $\phi(t)$ in the time segments $t\in[0, \tau_1]$ and $t\in[\tau_2, \tau]$ should be changed to $\xi_1-\pi/2$ and $\xi_2+\pi/2$, respectively. Notice that, the pulse shape here can be arbitrary, providing their pulse area are as prescribed. The final evolution operator is written as
\begin{eqnarray} \label{Utau}
U(\tau)&=&U(\tau,\tau_2)U(\tau_2,\tau_1)U(\tau_1,0)\\
&=&\left(
\begin{array}{cccc}
(c_{\gamma'} \!+\!\textrm{i}s_{\gamma'} c_{\chi_1}) e^{-\textrm{i} \xi_-}  & \textrm{i}s_{\gamma'} s_{\chi_1} e^{-\textrm{i} \xi_+} \\
\textrm{i}s_{\gamma'} s_{\chi_1} e^{\textrm{i} \xi_+} & (c_{\gamma'} \!-\!\textrm{i}s_{\gamma'} c_{\chi_1}) e^{\textrm{i} \xi_-}
\end{array}
\!\!\right),\notag
\end{eqnarray}
where $\gamma'=\gamma_g+\xi_-$ with the accumulated geometric phase in this evolution process being $\gamma_g=-(\xi_2-\xi_1)(1-\cos\chi_2)/2$. We find that geometric Hadamard gate $U^g_H$, Phase gate $U^g_S$ and $8/\pi$ gate $U^g_T$ can be realized by setting $\{\gamma'\!=\!\pi/4, \xi_2\!-\!\xi_1\!=\!3\pi, \chi_1\!=\!\pi/2\}$, $\{\gamma'\!=\!\pi, \xi_2\!-\!\xi_1\!=\!5\pi/2\}$ and $\{\gamma'\!=\!\pi, \xi_2\!-\!\xi_1\!=\!9\pi/4\}$, which constitute a universal set for single-qubit gates.

\begin{figure}[tbp]
  \centering
  \includegraphics[width=0.95\linewidth]{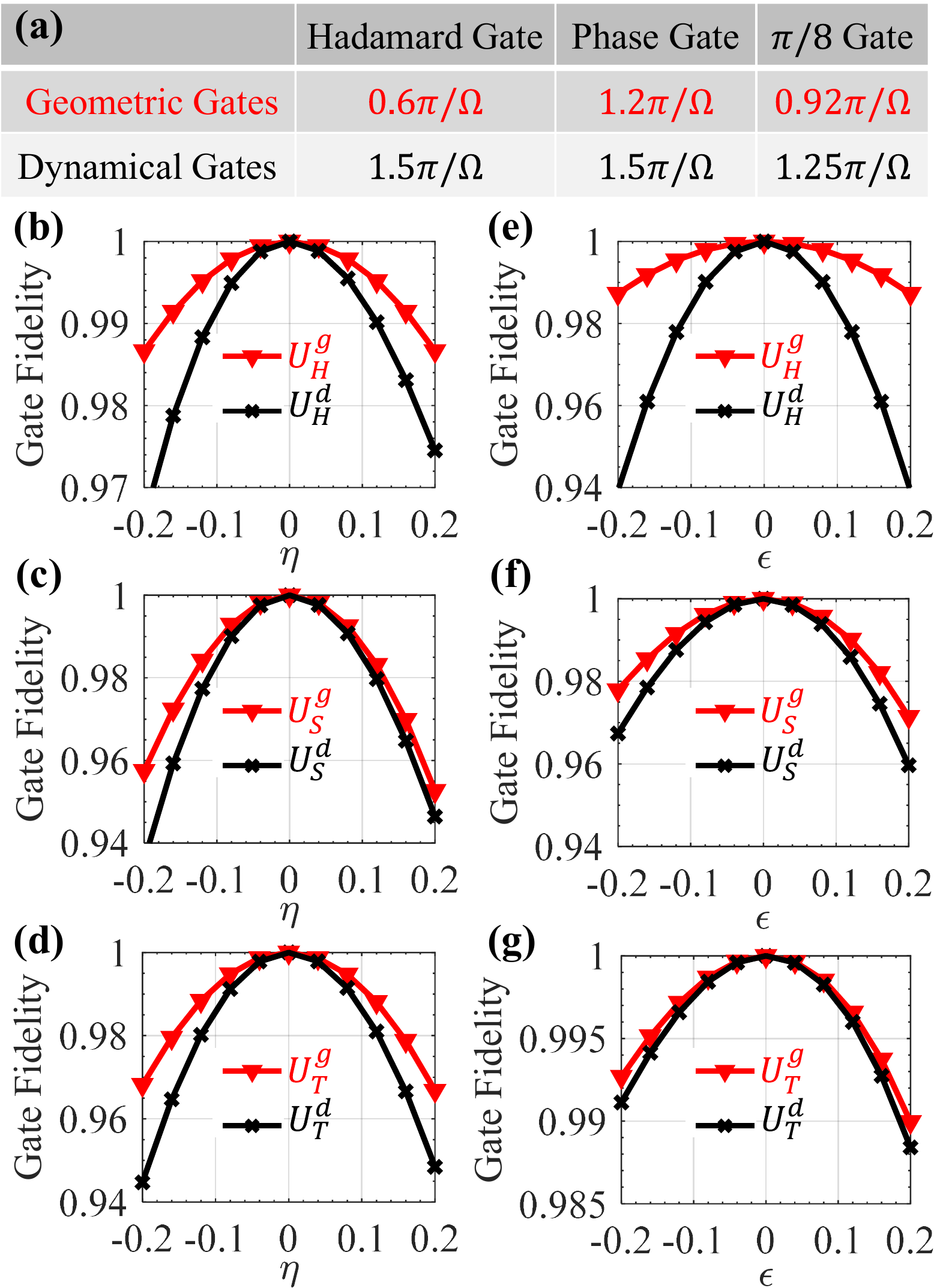}
  \caption{Comparison of our geometric gates and the conventional dynamical counterparts in terms of (a) gate time and (b)-(g) robustness.}
  \label{GateP}
\end{figure}

\section{The performance superiority of geometric gates}

Next, we continue to verify the advantages of our new geometric scheme over the conventional dynamical one (for details, see Appendix A), in terms of gate performance. Firstly, the needed gate-operation time is an important factor for gate performance, determines the impacts from environment-induced decoherence. From the comparison results are listed in Fig. \ref{GateP}(a), we can find that the time we consume to construct geometric gate is less than that of the dynamical counterpart, where $\Omega(t)=\Omega$ is set to be constant for simplicity. Secondly, we also consider the driving amplitude and detuning errors induced by the imperfect control, in the form of $(1+\epsilon)\Omega$ and $\Delta(t)+\eta\Omega$.
We use the formula $F_{\epsilon,\eta}=\textrm{Tr}(U^{\dagger}U_{\epsilon,\eta}) /\textrm{Tr}(U^{\dagger}U)$ to evaluate gate robustness in the presence of control errors, in which $U$ and $U_{\epsilon,\eta}$ are, respectively, the gate without and with error. The results of numerical simulation are shown in Figs. \ref{GateP}(b)-\ref{GateP}(g), which exhibit the advantage of our scheme in terms of both gate fidelity and robustness.

\begin{figure}[tbp]
  \centering
  \includegraphics[width=0.8\linewidth]{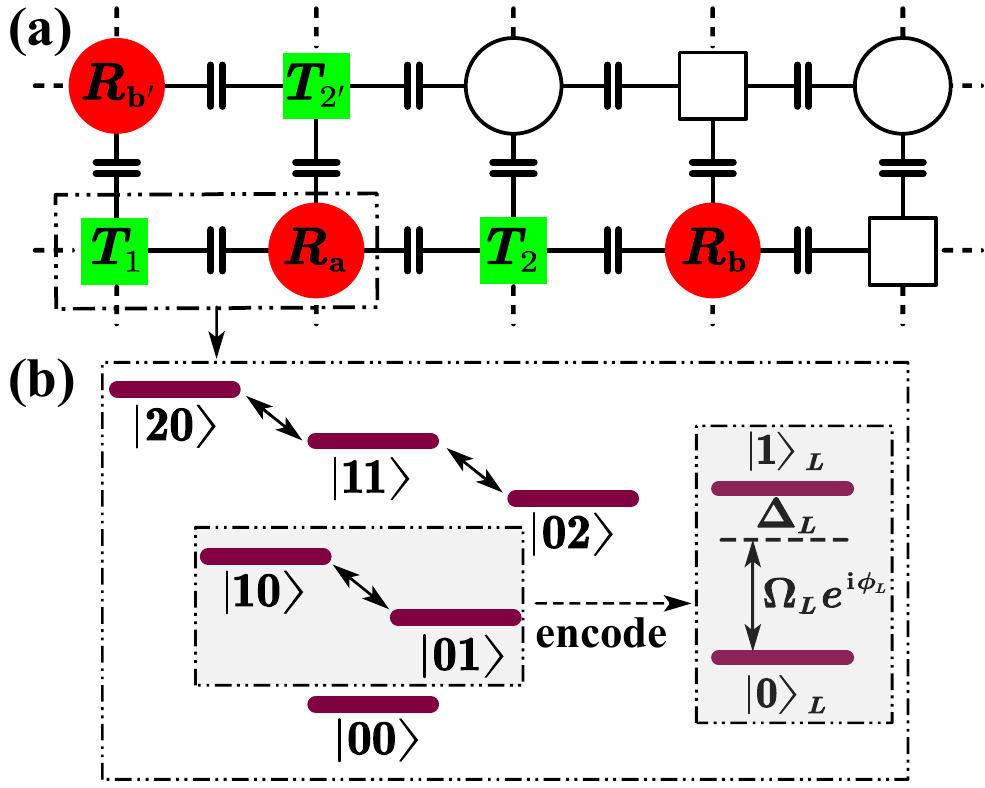}
  \caption{(a) A scalable 2D superconducting transmon-resonator lattice. (b) Energy level diagram of the capacitive coupling of a single transmon and a microwave resonator, where the single-excitation subspace $\{|10\rangle, |01\rangle\}$ as an effective two-level structure can encode a single-logical-qubit DFS.}
  \label{Set}
\end{figure}

\section{Implementation with encoding}

For the superconducting transmon qubit, the precise control of a single qubit, a common method is to apply the DRAG correction technology \cite{DRAGTho,DRAGExp} to suppress the computational basis leakage caused by the transmon's anharmonicity. However, it is necessary to set the drive to be time-dependent there, resulting in a longer gate time than the square pulse, so the decoherence effect and loss of gate robustness will also increase significantly. Therefore, we here propose to implement our universal geometric gates under the protection of DFS encoding on a two-dimensional (2D) transmon-resonator lattice as shown in Fig. \ref{Set}(a), which can combine error-tolerant features of our geometric scheme and decoherence resilience of DFS encoding. With the pursuit of a minimal number of physical qubits and without any additional auxiliary, we here only use capacitively coupled transmon $T_1$ and microwave resonator $R_a$ to encode a DFS logical qubit, thus there will exist a two-dimensional DFS $S_1=\{|10\rangle, |01\rangle\}$, where the encoded computational basis are denoted as $|0\rangle_L\!=\!|10\rangle$ and $|1\rangle_L\!=\!|01\rangle$ with $|mn\rangle=|m\rangle_{1}\otimes|n\rangle_{a}$.
For the parametrically tunable coupling \cite{TunCoupleTheo,TunCoupleExp} between a transmon $T_1$ and a fixed-frequency microwave resonator $R_a$, it is obtained by introducing an additional qubit-frequency driving for transmon $T_1$, in the form of $\omega_{q_1}(t)=\omega_{q_1}+\varepsilon_1\sin(\nu_1t+\varphi_1)$, which can be experimentally achieved by biasing the transmon with an ac magnetic flux. Energy level transition structure of the above interaction Hamiltonian as shown in Fig. \ref{Set}(b), there is naturally no leakage from DFS $S_1$ to the multi-excitation subspaces, thus no additional correction for the computational basis leakage is required. Furthermore, in the DFS representation $\{|0\rangle_L, |1\rangle_L\}$, the obtained effective Hamiltonian is of the same form as Eq. (\ref{H}) (for the derivation details, see Appendix B), thus arbitrary single-logical-qubit geometric gates, as in Eq. (\ref{Utau}), can be realized under the geometric parameter constraints in Eq. (\ref{3path}).

\begin{figure}[tbp]
  \centering
  \includegraphics[width=\linewidth]{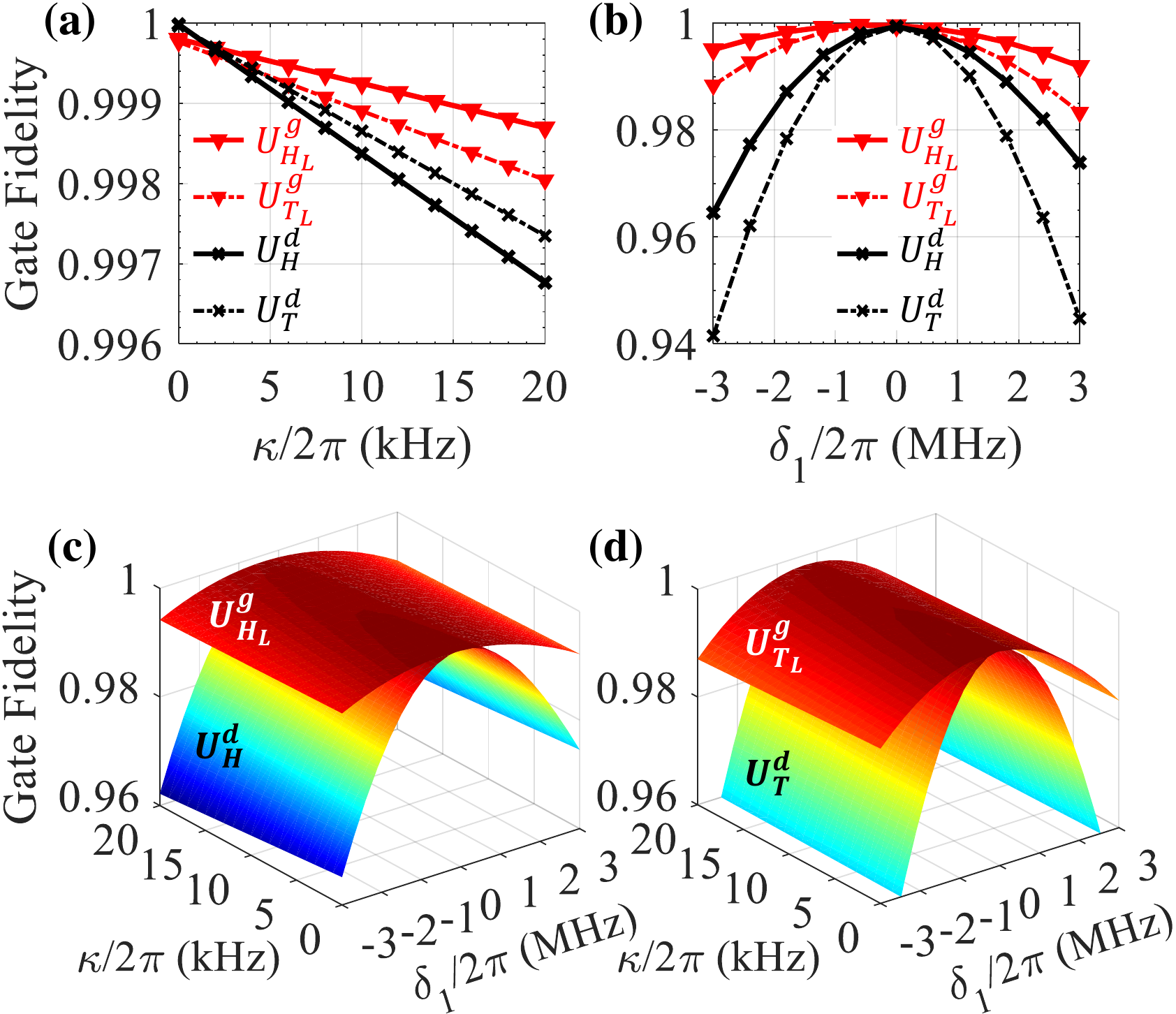}
  \caption{Under the effects of (a) decoherence and (b) qubit-frequency drift which are regard as the main source of gate errors in the superconducting quantum circuits, the performance comparison for our geometric gates under the protection of DFS encoding and the dynamical gates realized on a single transmon. Comparison of comprehensive effects for (c) Hadamard gate and (d) $\pi/8$ gate. Note that the performance of the Phase gate is similar to that of its same type of $\pi/8$ gate and thus not mention here and hereafter.}
  \label{GateP1}
\end{figure}

We take Hadamard gate $U^{g}_{H_L}$, Phase gate $U^{g}_{S_L}$ and $\pi/8$ gate $U^{g}_{T_L}$ as typical examples to numerically show the performance of these universal single-logical-qubit geometric gates by using the quantum master equation, where the simulation details are seen in Appendix D. According to the state-of-the-art technology in the experiments \cite{Dyn1,ResonatorEXP}, we here set decoherence rates $\kappa^1_-\!=\!\kappa^1_z\!=\!\kappa\!=\!2\pi\times4$ kHz, $\kappa_a\!=\!2\pi\times1$ kHz, and the coupling strength between transmon $T_1$ and microwave resonator $R_a$ to $g_{1a}\!=\!2\pi\!\times\!20$ MHz with anharmonicity of transmon $T_1$ as $\alpha_{1}\!=\!2\pi\!\times240$ MHz. At frequency difference between $T_1$ and $R_a$ as $\Delta_1=2\pi\times180$ MHz, the gate fidelities of geometric Hadamard gate, Phase gate and $\pi/8$ gate can be as high as 99.95\%, 99.92\% and 99.94\% by adjusting $\beta_1=\frac{\varepsilon_1}{\nu_1}\approx2.1$, respectively.
Besides, for superconducting quantum circuits, the frequency drift ($\omega_{q_1}\!\!\rightarrow\!\omega_{q_1}\!+\!\delta_1$) of a working transmon can up to a few of MHz. In Figs. \ref{GateP1}(a) and \ref{GateP1}(b), we plot the gate fidelities as functions of $\kappa$ and $\delta_1$, these results demonstrate that, although more physical resources are involved, our encoding scheme can still has a stronger robustness
than the conventional dynamical gates, which are realized by applying DRAG correction on a single transmon ($T_1$) under the same parameter settings. Remarkably, we consider both key errors, as shown in Figs. \ref{GateP1}(c) and \ref{GateP1}(d), which can further verify error-tolerant features of our scheme.

\begin{figure}[tbp]
  \centering
  \includegraphics[width=\linewidth]{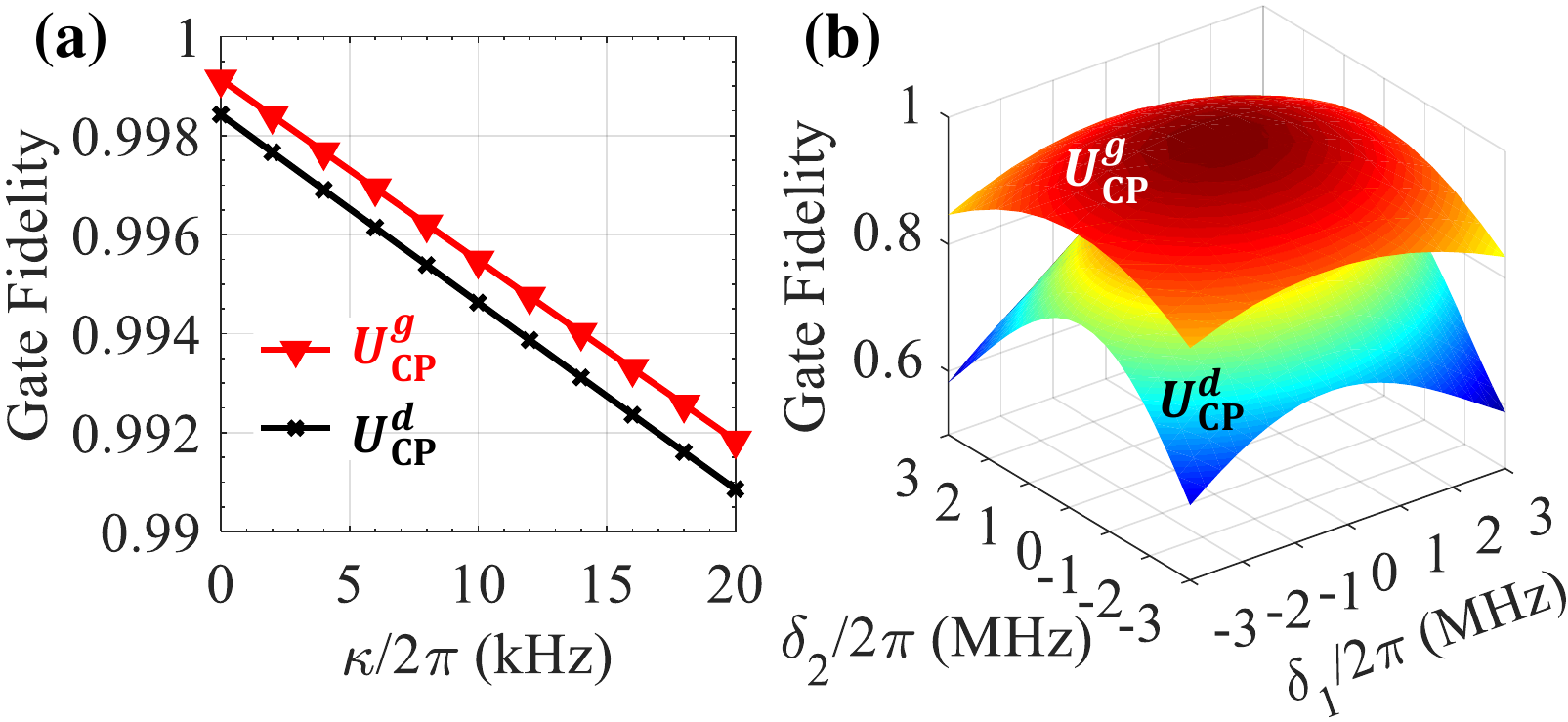}
  \caption{Under the effects of (a) decoherence and (b) qubit-frequency drifts ($\omega_{q_1}\!\!\rightarrow\!\omega_{q_1}\!+\delta_1$ and $\omega_{q_2}\!\!\rightarrow\!\omega_{q_2}\!+\delta_2$) of transmons $T_1$ and $T_2$, the performance comparison for two-logical-qubit controlled-phase geometric gate and two-physical-qubit dynamical counterpart.}
  \label{GateP2}
\end{figure}

Next, we extend our geometric approach to the situation of two logical qubits, as shown in Fig. \ref{Set}(a), and prove the overall gate-performance advantages compared to ``two-physical-qubit dynamical gate'' realized by tunable coupling between only two transmons. We continue to utilize a transmon $T_2$ and a microwave resonator $R_b$ on the same chain as the first logical qubit to encode the second logical qubit, thus there exists a four-dimensional DFS $S_2\!=\!\{|1010\rangle,\!|1001\rangle,\! |0110\rangle, \! |0101\rangle\}\!=\{|00\rangle_L,|01\rangle_L,|10\rangle_L, |11\rangle_L\}$ for two-logical qubits. In a same way, the derivation details
as seen in Appendix C, we can efficiently obtain two-logical-qubit controlled-phase gate
\begin{eqnarray} \label{EqU2}
U^g_{\textrm{CP}}(\zeta)=\textrm{diag}\{1,1,e^{\textrm{i}\zeta},1\},
\end{eqnarray}
with $\zeta$ being a geometric phase. In the following, we evaluate the performance of $U^g_{\textrm{CP}}(\pi/2)$ by utilizing the master equation, in which the effects of high-order oscillating terms and decoherence of transmons ($T_1$, $T_2$) and microwave resonators ($R_a$, $R_b$) are all taken into consideration. Here, we set the coupling strength $g_{{a2}}=\!2\pi\times 8$ MHz, and anharmonicity of the other transmon $T_2$ as $\alpha_2=2\pi\!\times 220$ MHz. When frequency difference between $T_2$ and $R_a$ as $\Delta_2=2\pi\times500$ MHz and driving parameter $\beta_2=\frac{\varepsilon_2}{\nu_2}\approx1.2$, we can simulate that the fidelity of two-logical-qubit controlled-phase geometric gate can up to 99.76\% under the decoherence rates $\kappa\!=\!2\pi\times 4$ kHz and $\kappa_a=\kappa_b=2\pi\times1$ kHz, which is comparable to the fidelity with 99.68\% of two-physical-qubit dynamical one $U^d_{\textrm{CP}}(\pi/2)$ realized only on transmons $T_1$ and $T_2$ under the same parameter settings. Furthermore, in Figs. \ref{GateP2}(a) and \ref{GateP2}(b), the numerical results indicate that our controlled-phase geometric gate shows a stronger error-tolerant feature over the two-physical-qubit dynamical gate.

\section{Conclusion}

In summary, we have proposed a fast and robust geometric scheme based on a simple two-level structure. And we merely utilize tunable coupling between a transmon and a microwave resonator to achieve high-fidelity geometric logical control in superconducting quantum circuits. Furthermore, from the numerical results, we can obviously find that, under hybrid protection of geometric phase and DFS encoding, our single- and tow-logical-qubit geometric gates can all show higher fidelity and greater robustness than the dynamical ones without DFS encoding. Therefore, all of these sufficiently demonstrate that our scheme paves the way for achieving the robust GQC.

\bigskip

\acknowledgments
The authors thank Z. Hua and S. Li for helpful discussions. This work was supported by the Key-Area Research and Development Program of GuangDong Province (Grant No. 2019B030330001), the CRF (No. C6009-20G)
of Hong Kong, the National Natural Science Foundation of China (Grant No. 11874156), and Science and Technology Program of Guangzhou (Grant No. 2019050001).

\appendix

\section{Dynamical quantum gates}

Conventional dynamical quantum gate \cite{Dyn1,Dyn2}, mentioned in the main text, can be realized by a resonant two-level drive. The corresponding Hamiltonian is as Eq. (\ref{H}) with $\Delta(t)=0$, in which driving phase $\phi(t)=\phi_d$ is a constant to ensure that there is no accumulation of geometric phase. In this way, dynamical Hadamard gate $U^d_H$, Phase gate $U^d_S$ and $8/\pi$ gate $U^d_T$ are, respectively, obtained by the operations of $R_x(\pi)R_y(\frac{\pi}{2})$, $R_y(-\frac{\pi}{2})R_x(\frac{\pi}{2})R_y(\frac{\pi}{2})$ and $R_y(-\frac{\pi}{2})R_x(\frac{\pi}{4})R_y(\frac{\pi}{2})$, where the corresponding operation elements $R_{x}(\theta)$ and $R_{y}(\theta)$ are the dynamical X- and Y-axis rotation operations for arbitrary angle $\theta\!=\!\int^{\tau}_0\!\Omega(t)\textrm{d}t$, which can be done by determining $\phi_d=0$ and $\pi/2$, respectively.

In addition, in the superconducting physical implementation of single-qubit dynamical gates, the corresponding control field needs to be corrected to $\textbf{B}_C(t)=\textbf{B}(t)+\textbf{B}_D(t)$, in which $\textbf{B}(t)$ and $\textbf{B}_D(t)\!=\!(-\dot{B}_y+\!B_zB_x, \dot{B}_x+B_zB_y, 0)/(2\alpha_1)$ are the original and additional DRAG-correcting microwave fields, respectively, where $\alpha_1$ is anharmonicity of the working transmon $T_1$. Consider the constraints of experimental control and DRAG correction on a single transmon, it is necessary to set a time-dependent pulse shape and ensure $\Omega(0)\!=\!\Omega(\tau)\!=\!0$. In the comparison of main text, we take $\Omega(t)=\Omega_{\textrm{m}}\sin(\pi t/\tau)$ as an example for simplicity.

\section{Tunable transmon-resonator coupling}


To achieve effective control for the single-logical-qubit states $|0\rangle_L$ ($|10\rangle$) and $|1\rangle_L$ ($|01\rangle$), we can utilize the parametrically tunable coupling between a transmon $T_1$ and a fixed-frequency microwave resonator $R_a$, which is obtained by introducing an additional qubit-frequency driving for transmon $T_1$. Move into the interaction picture, the system Hamiltonian can be written as
\begin{align} \label{EqHI}
\mathcal{H}^1_I(t)&= g_{_{1a}} \left\{ |10\rangle\langle 01|e^{\textrm{i}\Delta_1 t}e^{-\textrm{i}\beta_1 \cos(\nu_1 t+\varphi_1)} \right. \notag \\
&\left.\quad+\sqrt{2}|20\rangle\langle 11|e^{\textrm{i}(\Delta_1-\alpha_1) t}e^{-\textrm{i}\beta_1 \cos(\nu_1 t+\varphi_1)}\right. \notag \\
& \quad \left.+\sqrt{2}|11\rangle\langle 02|e^{\textrm{i}\Delta_1 t}e^{-\textrm{i}\beta_1 \cos(\nu_1 t+\varphi_1)}\right\} +\mathrm{H.c.},
\quad \tag{S1}
\end{align}
where $\beta_1=\varepsilon_1/\nu_1$; $g_{_{1a}}$ and $\Delta_1$ are coupling strength and the difference of the transition frequency between transmon $T_1$ and microwave resonator $R_a$; $\alpha_n$ is the intrinsic anharmonicity of transmon $T_n$. The energy level transition structure corresponding to the Hamiltonian $\mathcal{H}^1_I(t)$ as shown in Fig. \ref{Set}(b), we obviously find that the single-excitation subspace $\{|10\rangle, |01\rangle\}$ can form an effective two-level structure. Furthermore, when we encode the computational basis into DFS $S_1$, there is naturally no leakage from $S_1$ to the multi-excitation subspaces. Therefore, unlike the control in a single transmon, our encoding method is not limited by the transmon's anharmonicity, and also does not rely strictly on a time-dependent pulse shape for correcting the computational basis leakage.

We here utilize the Jacobi-Anger identity to expand e-index terms in Eq. (\ref{EqHI}), and modulate the qubit-driving frequency $\nu_1$ to satisfy $\Delta_1-\nu_1\!=\!-(\Delta_L\!+\mu)$ with $|\Delta_L+\mu|\ll\{\Delta_1, \nu_1\}$. Then, by neglecting the high-order oscillating terms, and applying the unitary transformation with the transformation matrix $U_R(t)\!=\!\textrm{exp}[-\textrm{i}(\Delta_L/2)\sigma^L_{z}t]$, the final effective Hamiltonian can be obtained as
\begin{align} \label{Heff}
\mathcal{H}_L(t)\!=\!\frac{1}{2}\Omega_L[\cos\phi_L(t)\sigma^L_x \!+\sin\phi_L(t)\sigma^L_y]\!-\!\frac{1}{2}\Delta_L\sigma^L_z, \tag{S2}
\end{align}

where $\sigma^L_{x,y,z}$ are the Pauli operators in the single-logical-qubit subspace $\{|0\rangle_L,\! |1\rangle_L\}$; the coupling strength $\Omega_L\!=\!2J_1(\beta_1)g_{1a}$ with $J_1(\beta_1)$ being a Bessel function of the first kind, and the relative phase $\phi_L(t)\!=\!\mu t+\varphi_1\!+\pi/2$. The above Hamiltonian form is equivalent to $\mathcal{H}(t)$ of Eq. (\ref{H}) under the DFS representation $\{|0\rangle_L, |1\rangle_L\}$.

\section{Two-logical-qubit geometric gate}

As shown in Fig. \ref{Set}(a), we continue to utilize another transmon $T_2$ and a resonator $R_b$ on the same chain as the first logical qubit to encode the second logical qubit. With the general scalability of our encoding scheme, we can also utilize $T_{2'}$ and $R_{b'}$ in the column direction of two-dimensional superconducting lattice to encode the second logical qubit. In addition, for the arbitrary target control of the two-logical-qubit states, we can realize it by the parametrically tunable coupling between adjacent microwave resonator $R_a$ and transmon $T_2$,  in which transmon $T_2$ is driven by an additional qubit-frequency driving in the form of $\omega_{q_2}(t)=\omega_{q_2}+\varepsilon_2\sin(\nu_2t+\varphi_2)$. Thus, the corresponding interaction Hamiltonian can be described as
\begin{align} \label{EqHT}
\mathcal{H}^2_I(t)&= g_{_{a2}} \left\{ |01\rangle_{a2}\langle 10|e^{\textrm{i}\Delta_2 t}e^{-\textrm{i}\beta_2 \cos(\nu_2 t+\varphi_2)} \right. \notag \\
&\left.\quad+\sqrt{2}|02\rangle_{a2}\langle 11|e^{\textrm{i}(\Delta_2-\alpha_2) t} e^{-\textrm{i}\beta_2 \cos(\nu_2 t+\varphi_2)}\right. \notag \\
& \quad \left. +\sqrt{2}|11\rangle_{a2}\langle 20|e^{\textrm{i}\Delta_2 t}e^{-\textrm{i}\beta_2 \cos(\nu_2 t+\varphi_2)}\right\} +\mathrm{H.c.},\tag{S3}
\end{align}
where $\beta_2=\varepsilon_2/\nu_2$; $g_{_{a2}}$ and $\Delta_2$ are coupling strength and the difference of the transition frequency between transmon $T_2$ and microwave resonator $R_a$. Energy level transition structure of the Hamiltonian $\mathcal{H}^2_I(t)$ is similar to Fig. \ref{Set}(b). But, different from the single-logical-qubit case manipulated in the single-excitation subspace, the manipulation of our two-logical-qubit states need to be implemented in the two-excitation subspace $\{|02\rangle_{a2}, |11\rangle_{a2}\}$ to realize a nontrivial controlled-phase gate. To this end, we here modulate driving frequency $\nu_2$ to satisfy $\Delta_2-\alpha_2-\nu_2=\!-(\Delta_{L_2}+\upsilon)$ with $|\Delta_{L_2}+\upsilon|\ll\{\Delta_2-\alpha_3, \nu_2\}$, and apply the unitary transformation, the Hamiltonian mapped into the DFS $S_2$ representation can then read as
\begin{align} \label{EqHL2}
\mathcal{H}_{L_2}(t)&=-\frac{\Delta_{L_2}}{2}(|a\rangle_L\langle a|-|11\rangle_L\langle 11|) \notag \\
&\quad +\frac{\Omega_{L_2}}{2}(e^{-\textrm{i}\phi_{L_2}(t)}|a\rangle_L\langle 11|+\mathrm{H.c.}),\tag{S4}
\end{align}
where the auxiliary state $|a\rangle_L\!=\!|0020\rangle$, and the effective coupling strength and relative phase are $\Omega_{L_2}\!\!=\!2\sqrt{2}J_1(\beta_2)g_{a2}$ and $\phi_{L_2}(t)\!=\!\upsilon t\!+\!\varphi_2\!+\!\pi/2$, respectively. For a set of orthogonal evolution states
\begin{eqnarray} \label{Evostate1}
|\Psi^L_{0}(t)\rangle &=&e^{\textrm{i}F_0(t)}[\cos{\frac {\chi^{_L}(t)} {2}}|a\rangle_L+\sin{\frac {\chi^{_L}\!(t)} {2}}e^{\textrm{i}\xi^{L}(t)}|11\rangle_L], \notag\\
|\Psi^L_{1}(t)\rangle &=&e^{\textrm{i}F_1(t)}[\sin{\frac {\chi^{_L}(t)} {2}}e^{-\textrm{i}\xi^{L}(t)}|a\rangle_L-\cos{\frac {\chi^{_L}(t)} {2}}|11\rangle_L], \notag
\end{eqnarray}
under the control of $\mathcal{H}_{L_2}(t)$, similar to derivation process of Eq. (\ref{paralimt1}) to Eq. (\ref{3path}), in addition to meeting geometric condition, we also need to determine parameter $\chi^{_L}_1=0$ to ensure that the evolution state $|\Psi^L_{1}(t)\rangle$ will not leak to the non-computational subspace at the final moment $t\!=\!\tau'$, and accumulate a geometric phase as $\gamma_g\!=\!-\xi^{L}_2 (1\!-\cos\chi^{_L}_2)/2$ in the two-logical-qubit state $|11\rangle_L$, where defining $\xi^{L}_1=0$ for simplicity. Therefore, by engineering the Hamiltonian parameters $\Omega_{L_2}$ and $\phi_{L_2}\!(t)$ in the time segments $t\in[0, \tau'_1]$, $[\tau'_1, \tau'_2]$ and $[\tau'_2, \tau']$ as
\begin{align} \label{3path2}
\Omega_{L_2} \tau'_1&=\chi^{_L}_2, \quad\quad\quad\quad \ \ \phi_{L_2}(t)=\frac{\pi}{2}, \notag\\
\Omega_{L_2} (\tau'_2-\tau'_1)&=2\xi^{L}_2\sin(2\chi^{_L}_2), \ \ \!\!\phi_{L_2}(t)=\pi+\frac{\Omega_{L_2}(t-\tau'_1)}{2\sin(2\chi^{_L}_2)}, \notag\\
\Omega_{L_2} (\tau'-\tau'_2)&=\chi^{_L}_2,\quad\quad\quad\quad \ \ \phi_{L_2}(t)=\xi^{L}_2-\frac{\pi}{2},\tag{S5}
\end{align}
respectively, with $\Delta_{L_2}\!=\!0, -\Omega_{L_2}\!\tan\!\chi^{L}_2, 0$, the final evolution operator within two-logical-qubit DFS $S_2$ is
\begin{align} \label{EqU2}
U^g_{\textrm{CP}}(\zeta)=\textrm{diag}\{1,1,e^{\textrm{i}\zeta},1\}, \tag{S6}
\end{align}
in which $\zeta\!\!=\!-\gamma_g\!=\!\xi^{L}_2 (1\!-\!\cos\chi^{_L}_2)/2$. Notice that, for the two-logical-qubit controlled-phase geometric gate $U^g_{\textrm{CP}}(\pi/2)$ with the target phase $\zeta\!=\pi/2$, parameter $\chi^{_L}_2$ (or $\xi^{_L}_2$) still has different choices to realized a same gate type at the final time $\tau'=2[\chi^{_L}_2+\xi^{L}_2\sin(2\chi^{_L}_2)]/\Omega_{L_2}$, so we determine $\chi^{_L}_2=0.56\pi$ to gain the least gate-time consumption.

\section{Quantum master equation and calculating fidelity}

Quantum system is inevitably affected by decoherence due to its coupling with the surrounding environment, so we need to consider the effects of decoherence and the neglected high-order oscillating terms simultaneously in our subsequent numerical simulation. The performance of our universal logical-qubit gates can be numerically evaluated by using the quantum master equation as follows:
\begin{align} \label{EqMaster}
\dot\rho_n\!=\!-\textrm{i}[\mathcal{H}^n_I(t), \rho_n]\!+\!\!\!\sum_{n'=1}^n\sum_{u=-,z}\!\! \frac {\kappa^{n'}_u} {2}\mathscr{L}(X^{n'}_{u})\!+\!\!\!\sum_{w=a,b}\!\!\! \frac {\kappa_w} {2}\mathscr{L}(Y_{w}), \tag{S7}
\end{align}
where $\rho_{n}$ is the density matrix of quantum system under consideration, $\mathscr{L}(\mathcal{A})=2\mathcal{A}\rho_{_n}
\mathcal{A}^\dagger-\mathcal{A}^\dagger \mathcal{A} \rho_{_n} -\rho_{_n} \mathcal{A}^\dagger \mathcal{A}$ is the Lindblad operator for operator $\mathcal{A}$ with $X^n_-\!=\!\!\sum_{j=0}^{+\infty}\!\sqrt{j+1}|j\rangle_{n} \langle j\!+\!1|$, $X^n_z\!=\!\sum_{j=0}^{+\infty}j|j\rangle_{n} \langle j|$ and $Y_w=\sum_{j=0}^{+\infty}|j\rangle_{w} \langle j+1|$; $\kappa^n_-$ and $\kappa^n_z$ are the decay and dephasing rates of transmon $T_n$, and $\kappa_w$ is the decay rate of microwave resonator $R_w$. $n\!=\!1$ and $n\!=\!2$ represent the single- and two-logical-qubit cases, respectively. Notice that, for the case of two-logical qubits, we consider the effects of the high-order oscillating terms and decoherence of all transmons ($T_1$, $T_2$) and microwave resonators ($R_a$, $R_b$). Therefore, by numerically solving the above quantum master equation, we can obtain the final density matrices $\rho_{_{f1}}$ and $\rho_{_{f2}}$ of single qubit and two-logical qubits.

We next use the solved density matrix to fully evaluate our implemented geometric quantum gates based on DFS encoding. Thus we can define the gate fidelity of single-logical qubit as: $F_1^G=\frac {1} {2\pi}\int_0^{2\pi} \langle \psi_{f_1}|\rho_{_{f1}}|\psi_{f_1}\rangle \textrm{d}\theta_1$, where $|\psi_{f_{1}}\rangle\!=\!U(\tau)|\psi_1\rangle$ is the ideal final state for general initial state of single-logical qubit $|\psi_1\rangle=\cos\theta_1|0\rangle_L+\sin\theta_1|1\rangle_L$. In the same way, based on solved density matrix $\rho_{_{f2}}$, two-logical-qubit gate fidelity can be defined as: $F^G_{\textrm{CP}}=\frac {1} {4\pi^2}\int_0^{2\pi} \int_0^{2\pi} \langle \psi_{f_{\textrm{CP}}}|\rho_{_{f2}}|\psi_{f_{\textrm{CP}}}\rangle d\vartheta_1d\vartheta_2$ where $|\psi_{f_{\textrm{CP}}}\rangle\!=\!U^g_{\textrm{CP}}(\pi/2)|\psi_2\rangle$ is the ideal final state for general initial state of two-logical qubits  $|\psi_2\rangle\!=\!(\cos\vartheta_1|0\rangle_L\!+\sin\vartheta_1|1\rangle_L) \otimes(\cos\vartheta_2|0\rangle_L+\sin\vartheta_2|1\rangle_L)$.

\end{document}